\documentclass[12pt]{article}

\usepackage{cite}

\newcommand{\pd}[2]{\frac{\partial #1}{\partial #2}}
\newcommand{\asinh}{\textrm{arcsinh}}

\title{The WKB approximation in the deformed space
with the minimal length and minimal momentum}
\author{
T.V. Fityo\footnote{E-mail: fityo@ktf.franko.lviv.ua},\ \ I.O.
Vakarchuk\footnote{E-mail: chair@ktf.franko.lviv.ua}\ \ and V.M.
Tkachuk\footnote{E-mail: tkachuk@ktf.franko.lviv.ua}
\\
  {\small Chair of Theoretical Physics, Ivan Franko National University of Lviv,}\\
{\small 12 Drahomanov St., Lviv, UA-79005, Ukraine
         }}

\begin{document}

\maketitle

\abstract{A Bohr-Sommerfeld quantization rule is generalized for
the case of the deformed commutation relation leading to minimal
uncertainties in both coordinate and momentum operators. The
correctness of the rule is verified by comparing obtained results
with exact expressions for corresponding spectra.}



\section{Introduction}

Several independent lines of theoretical physics investigations
(e.g. string theory and quantum gravity) suggest the existence of
finite lower bound to the possible resolution of length $\Delta X$
(minimal length) \cite{Gross88, Maggiore93, Witten96}. Such an
effect can be achieved by modifying the usual canonical
commutation relations \cite{Kempf94, Kempf95,
Maggiore94,Detournay02}.

In our previous article \cite{Fityo06}, we have developed a
semiclassical approach for the deformed space with the following
commutation relation between the coordinate $\hat X$ and the
momentum $\hat P$ operators:
\begin{eqnarray}\label{s0}
[\hat X,\hat P]=i\hbar f(\hat P),
\end{eqnarray}
where $f$ is a positive function of the momentum. Such a
deformation may lead to the existence of the minimal length. Using
the semiclassical analysis, we have shown that in the deformed
space usual Bohr-Sommerfeld quantization rule for canonically
conjugated variables, $x$ and $p$,
\begin{eqnarray}\label{s01}
\oint pdx=2\pi\hbar(n+\delta)
\end{eqnarray}
is valid \cite{Fityo06}. This expression was successfully used to
consider a one-dimensional Coulomb-like problem \cite{Fityo06b}.
Parameter $\delta$ depends on boundary conditions: for smooth
potentials and deformation function $f(P)$ such that $f(0)\ne0$,
it equals $1/2$. The rule (\ref{s01}) can be rewritten in terms of
classical variables $X$ and $P$ corresponding to the initial
coordinate and momentum operators:
\begin{eqnarray}\label{s02}
-\oint \frac{XdP}{f(P)}=2\pi\hbar(n+\delta).
\end{eqnarray}

\section{Bohr-Sommerfeld quantization rule}

The purpose of this paper is to generalize result (\ref{s02}) for
a wider class of the deformed commutation relation
\begin{eqnarray}\label{s1}
[\hat X,\hat P]=i\hbar f(\hat X,\hat P),
\end{eqnarray}
where $f$ is a positive function of the position and momentum.
Such a modification of the canonical commutation relation can lead
to the existence of nonzero minimal uncertainties of the position
and the momentum operators. Function,
\begin{equation}\label{s1a}
f(\hat X,\hat P)=1+\alpha\hat X^2+\beta\hat P^2,
\end{equation}
is an example of the deformation leading to the minimal length and
minimal momentum. In this case $\Delta
X=\hbar\sqrt{\beta/(1-\hbar^2\alpha\beta)}$, $\Delta
P=\hbar\sqrt{\alpha/(1-\hbar^2\alpha\beta)}$. We use only  this
deformation function in this paper, considering examples in the
third section.

Let us know the representation of operators $\hat X$ and $\hat P$
satisfying (\ref{s1})
\begin{equation}\label{s1b}
\hat X=X(\hat x,\hat p),\qquad \hat P=P(\hat x,\hat p),
\end{equation}
where small operators $\hat x$, $\hat p$ are canonically
conjugated. It seems that this representation can be found
formally as series over deformation parameters (for examples of
such expansion see \cite{Quesne07}). The explicit representation
of $\hat X$ and $\hat P$ operators for deformation function
(\ref{s1a}) can be found in \cite{Quesne03}. Note that when the
deformation function depends only on the position, i. e. $f=f(X)$,
the commutation relation describes a particle with a
position-dependent mass $M(X)=1/f^2(X)$ \cite{Quesne04p}.

In the classical limit $\hbar\to0$ the deformed commutation
relation (\ref{s1}) leads to the deformed Poisson bracket
\cite{Benczik02,Frydryzhak03}
\begin{eqnarray}\label{s2m}
\{X,P\}=f(X,P).
\end{eqnarray}
It is always possible to choose such canonically conjugated
variables $x$ and $p$  (Darboux theorem \cite{Arnold89}) that $X$
and $P$ as functions of $x$, $p$ satisfy
\begin{eqnarray}\label{s3}
\{X,P\}_{x,p}=\pd{X}{x}\pd Pp-\pd Px\pd Xp= f(X,P).
\end{eqnarray}

We consider that $P(x,p)$ is an odd monotonic function with
respect to $p$ and even function of $x$; $X(x,p)$ is an odd
monotonic function with respect to $x$ and an even function of
$p$. This provides left-right symmetry and one-to-one
correspondence between $P$ and $p$, $X$ and $x$, respectively. The
one-dimensional Schr\"odinger equation in the deformed space reads
\begin{equation}\label{s00}
\frac{\hat P^2}{2m}\psi+U(\hat X)\psi=E\psi.
\end{equation}
Let us write the wave function in the following form
$$\psi(x)=\exp\left[\frac{i}{\hbar}S(x) \right]$$
then in the linear approximation over $\hbar$
\begin{equation}\label{bs1}
H\left(x,-i\hbar\frac d{dx}\right)\psi(x)=\left[
H(x,S'(x))-\frac{i\hbar}2 H_{pp}(x,S'(x))S''(x)+\dots
\right]\psi(x),
\end{equation}
where subscript $p$ denotes derivative with respect to the second
argument of $H(x,p)$ and the prime denotes derivative with respect
to $x$. Formula (\ref{bs1}) is derived for a normally ordered
Hamiltonian (the powers of operator $\displaystyle
-i\hbar\frac{d}{dx}$ act on the wavefunction first and then are
multiplied by the functions of $x$). Expanding $S(x)$ in power
series over $\hbar$
\begin{equation}\label{bs2}
S(x)=S_0(x)+\frac{\hbar}iS_1(x)+\dots
\end{equation}
from (\ref{bs1}) we obtain the following set of equations for $S$:
\begin{eqnarray}
H\left(x,S'_0(x)\right)=E,\label{bs3}\\
H_p\left(x,S_0'(x)\right)\frac\hbar iS_1'(x)-\frac{i\hbar}2
H_{pp}\left(x,S_0'(x)\right)S''_0(x)=0.\label{bs4}
\end{eqnarray}
From the requirement that $P(x,p)$ is an odd monotonic function
with respect to $p$ and the Hamiltonian form $\displaystyle
H=\frac{P^2}{2m}+ U(X)$ we obtain that equation (\ref{bs3}) has
only two solutions with respect to $S_0'(x)$. They differs only by
sign, $S'_0(x)=\pm p(x)$. Then equation (\ref{bs4}) can be
rewritten as follows
$$H_p(x,\pm p)S_1'(x)=\mp\frac12p'H_{pp}(x,\pm p)$$
and $S_1(x)$ can be found from it. Note, that since $H(x,p)$ is an
even function with respect to $p$, $S_1$ is also an even function
of $p$ and thus does not depend on $S_0(x)$ sign. Then the
wavefunction in the linear approximation over $\hbar$ is
\begin{equation}\label{bs5}
\psi(x)=\exp{\left[S_1(x)\right]} \left( C_1
\exp\left[\frac{i}{\hbar} \int^xpdx\right]+
 C_2 \exp\left[ -\frac{i}{\hbar} \int^xpdx \right]\right).
\end{equation}
To obtain an expression of the Bohr-Sommerfeld quantization rule,
we have to analyze the behavior of wavefunction (\ref{bs5}) at the
infinities and consider matching conditions near the turning
points. The obtained expression (\ref{bs5}) is similar to the
corresponding expression of non-deformed quantum mechanics
\cite{Landau}. Performing the analogous analysis, we obtain for
bound states the same expression for the Bohr-Sommerfeld
quantization condition
\begin{equation}\label{h12}
\int\limits^{x_2}_{x_1} pdx = \pi\hbar(n+\delta), \quad
n=0,1,2,\dots
\end{equation}
where $x_1$ and $x_2$ are the turning points satisfying equation
$U(x)=E$, $\delta$ depends on boundary conditions and properties
of $S_1(x)$. But it is simpler to find the value of $\delta$ from
the limiting procedure to the non-deformed case. Such a procedure
is illustrated with the help of examples. We also would like to
recall that for large quantum numbers exact $\delta$ value is not
significant.

Condition (\ref{h12}) has the same form as in non-deformed quantum
mechanics only for auxiliary canonically conjugated variables. It
is possible to rewrite it in the initials variables $X$ and $P$.
Rewriting expression (\ref{h12}) in more convenient form
\begin{eqnarray}\label{s5}
\int\limits_{H\le E_n} dpdx=2\pi\hbar(n+\delta)
\end{eqnarray}
and noting that the Jacobian of changing from $x$ and $p$
variables to the initial variables $X$ and $P$ has the form
$\displaystyle
\pd{(x,p)}{(X,P)}=\frac1{\pd{(X,P)}{(x,p)}}=\frac1{f(X,P)}$, we
can conclude that
\begin{eqnarray}\label{s7}
\int\limits_{H(P,X)\le E_n}
\frac{dXdP}{f(X,P)}=2\pi\hbar(n+\delta).
\end{eqnarray}

\section{Examples}

In this section, we consider two eigenvalue problems, namely, the
harmonic oscillator and the potential well. We analyze them for
the following deformation of the canonical commutation relation
\begin{eqnarray}\label{s11}
[\hat X,\hat P]=i(1+\alpha \hat X^2+\beta\hat P^2).
\end{eqnarray}
Such a deformation was considered for the first time in
\cite{Kempf94} and it leads to the existence of nonzero minimal
uncertainties of the position and the momentum operators.

Let us consider the harmonic oscillator eigenvalue problem
\begin{eqnarray}\label{s10}
(\hat P^2+\hat X^2)\psi=E\psi.
\end{eqnarray}
The Bohr-Sommerfeld quantization rule (\ref{s7}) gives
\begin{eqnarray}\label{s12}
E_n=\frac{(\sqrt\alpha+\sqrt\beta)^2}{4\alpha\beta}
e^{2(n+\delta)\sqrt{\alpha\beta}}+\frac{(\sqrt\alpha-\sqrt\beta)^2}{4\alpha\beta}
e^{-2(n+\delta)\sqrt{\alpha\beta}}
-\frac{\alpha+\beta}{2\alpha\beta}.
\end{eqnarray}
Let us compare this result with exact one obtained in
\cite{Quesne03} which can be written in the following form
\begin{eqnarray}\label{s13}
E_n=a q^n + b + c q^{-n},
\end{eqnarray}
where $q=\frac{1+\sqrt{\alpha\beta}}{1-\sqrt{\alpha\beta}} \approx
e^{2\sqrt{\alpha\beta}}$; $a$, $b$, $c$ are coefficients depending
on the deformation parameters $\alpha$ and $\beta$ in a certain
cumbersome way. For small $\alpha$ and $\beta$, these coefficients
can be approximated as follows:
\begin{eqnarray}\label{s14}
a=\frac{(\sqrt\alpha+\sqrt\beta)^2}{4\alpha\beta}
\left(1+\sqrt{\alpha\beta}\right)+o(1),\nonumber\\
b=-\frac{\alpha+\beta}{2\alpha\beta}+o(1),\\
c=\frac{(\sqrt\alpha-\sqrt\beta)^2}{4\alpha\beta}
\left(1-\sqrt{\alpha\beta}\right)+o(1).\nonumber
\end{eqnarray}
The leading terms of $a$, $b$ and $c$ coincide with the
corresponding coefficients in expression (\ref{s12}). We would
like to stress that the Bohr-Sommerfeld quantization rule
(\ref{s7}) provides the exponential dependence of the spectrum on
the quantum number, $n$, as it should be according to the exact
result (\ref{s13}).

In linear approximation over deformation parameters, the WKB
approximation (\ref{s12}) for $\delta=\frac12$ (we choose this
value of $\delta$ to obtain an undeformed spectra $E_n=2n+1$ if
$\alpha=\beta=0$) gives
\begin{eqnarray}\label{s15}
E_n=2n+1+(\alpha+\beta)\left(n+\frac12\right)^2.
\end{eqnarray}
While the exact expression for the spectrum differs from it by
$\frac14(\alpha+\beta) + o(\alpha,\beta)$.

The second example is the potential well described by Hamiltonian
\begin{eqnarray}\label{s16}
\hat H=\hat P^2,
\end{eqnarray}
where the coordinate $X$ is considered to be bounded in the
interval $[-a,a]$. Then Bohr-Sommerfeld quantization rule reads
\begin{eqnarray}\label{s17}
\int\limits^a_{-a}\int\limits^{\sqrt {E_n}}_{-\sqrt{E_n}}\frac{dX
dP}{1+\alpha X^2 + \beta P^2}=2\pi(n+\delta).
\end{eqnarray}
We failed to integrate this expression exactly, thus we restrict
ourselves to consider only the linear approximation over
deformation parameters. For energy levels we obtain
\begin{eqnarray}\label{s18}
E_n= \left(\frac{\pi n}{2a}\right)^2\left[ 1+\frac23 \alpha a^2
+\frac23\beta\left(\frac{\pi n}{2a}\right)^2
\right]+o(\alpha,\beta).
\end{eqnarray}
Here we fixed the value of $\delta$ as $0$ to reproduce the
spectrum of the potential well in the undeformed case
($\alpha=\beta=0$).

The Bohr-Sommerfeld quantization rule (\ref{s7}) predicts that the
potential well (\ref{s16}) has only a finite amount of bound
states. To show this let us consider the case of large energies
$E_n\to\infty$. Since%
$$
\int\limits^a_{-a}\int\limits^{\infty}_{-\infty}\frac{dX
dP}{1+\alpha X^2 + \beta P^2}=\frac{2\pi}
{\sqrt{\alpha\beta}}\asinh \sqrt\alpha a$$%
is finite, equation (\ref{s17}) has no solution for sufficiently
large $n$. The maximal value of $n$ for which it can be solved
$n_{max}=[\asinh \sqrt\alpha a/\sqrt{\alpha\beta}]$, where
$[\dots]$ denotes the integer part. In limit $\alpha\to 0$ we
obtain $n_{max}=[a/\sqrt\beta]$. This number coincides with the
amount of bound states obtained in \cite{Detournay02} for the
potential well with $\alpha=0$. For $\alpha=0$ the eigenvalue
problem was considered in \cite{Fityo06,Brout99,Detournay02} and
expression (\ref{s18}) agrees with the results of these papers.

In the case of $\beta=0$, integral (\ref{s17}) can be calculated
exactly and
\begin{eqnarray}\label{s19}
E_n= \left(\frac{\pi n\sqrt\alpha}{2\arctan(\sqrt\alpha a)
}\right)^2.
\end{eqnarray}
An eigenvalue problem corresponding to Hamiltonian (\ref{s16}) and
the case $\beta=0$ can be easily solved exactly and the spectrum
coincides with expression (\ref{s19}). It is interesting to note
that in the limit $a\to\infty$, we effectively obtain a free
particle with a position-dependent mass and discrete spectrum for
it: $E_n=\alpha n^2$.

\section{Conclusions}

In the paper, we consider the Bohr-Sommerfeld quantization rule
for the case when the right hand of commutation relation depends
both on the coordinate and the momentum operators. To validate the
obtained Bohr-Sommerfeld quantization rule, we analyze two
examples with its help and compare the obtained results with
already known ones.

The case of exactly solvable harmonic oscillator with the
deformation $f(\hat X,\hat P)=1+\alpha\hat X^2+\beta\hat P^2$
shows that the Bohr-Sommerfeld quantization rule is applicable for
small values of the deformation parameters $\alpha$, $\beta$. The
second example is related to the potential well problem which has
not been solved exactly in the deformed case for arbitrary
$\alpha$ and $\beta$. The spectrum obtained with the help of
Bohr-Sommerfeld quantization rule agrees either with the already
known spectrum for $\alpha=0$ or for easily evaluated spectrum
with $\beta=0$.

This agreement is a good reason to consider that the
Bohr-Som\-mer\-feld quantization rule (\ref{s7}) is valid for
small values of the deformation parameters and can be used as an
approximate method for the spectrum estimation.


\begin{thebibliography}{99}

\bibitem{Gross88} D.~J.~Gross and P.~F.~Mende, Nucl. Phys. B
{\bf 303}, 407 (1988).
\bibitem{Maggiore93} M.~Maggiore, Phys. Lett. B {\bf 304}, 65
(1993).
\bibitem{Witten96} E.~Witten, Phys. Today {\bf 49}, 24 (1996).

\bibitem{Kempf94} A.~Kempf, J. Math. Phys. {\bf 35}, 4483  (1994).
\bibitem{Kempf95} A.~Kempf, G.~Mangano and R.~B.~Mann,
Phys. Rev. D {\bf 52}, 1108 (1995).
\bibitem{Maggiore94} M.~Maggiore, Phys. Rev. D {\bf 49}, 5182
(1994).
\bibitem{Detournay02} S.~Detournay, C.~Gabriel and Ph.~Spindel,
Phys. Rev. D {\bf 66}, 125004 (2002).

\bibitem{Fityo06} T.~V.~Fityo, I.~O.~Vakarchuk and
V.~M.~Tkachuk, J. Phys. A {\bf 39}, 379 (2006).

\bibitem{Fityo06b} T.~V.~Fityo, I.~O.~Vakarchuk and
V.~M.~Tkachuk, J. Phys. A {\bf 39}, 2143 (2006).

\bibitem{Quesne07} C.~Quesne and V.~M.~Tkachuk, SIGMA {\bf 3}, 016
(2007).

\bibitem{Quesne03} C.~Quesne and V.~M.~Tkachuk,  J. Phys. A
{\bf 36}, 10373 (2003).

\bibitem{Quesne04p} C.~Quesne and V.~M.~Tkachuk
J. Phys. A {\bf 37}, 4267 (2004).

\bibitem{Benczik02} S.~Benczik, L.~N.~Chang, D.~Minic,
N.~Okamura, S.~Rayyan and T.~Takeuchi, Phys. Rev. D {\bf 66},
026003 (2002).

\bibitem{Frydryzhak03} A.~M.~Frydryzhak and V.~M.~Tkachuk,
Chechoslovak J. Phys. {\bf 53}, 1035 (2003).

\bibitem{Arnold89} V.~I.~Arnold {\em Mathematical Methods
of Classical Mechanics}, Springer-Verlag: New York (1989).

\bibitem{Landau} L.~D.~Landau, E.~M.~Lifshits, {\it Quantum mechanics:
non-relativistic theory} Pergamon Press: New York (1977).

\bibitem{Brout99} R.~Brout, C.~Gabriel, M.~Lubo and Ph.~Spindel,
Phys. Rev. D {\bf 59}, 044005 (1999).

\end{thebibliography}
\end{document}